\documentclass[a4paper,11pt]{article}

\usepackage{jheppub} 

\usepackage[T1]{fontenc} 
\usepackage{xcolor}
\usepackage{amssymb}
\usepackage{amsmath}
\usepackage{cases}
\usepackage{epsfig,subfigure}
\usepackage{graphicx}
\usepackage{epstopdf}
\usepackage{soul}
\usepackage{placeins}
\usepackage{CJK}
\usepackage{color}
\usepackage{slashed}

\newcommand {\nn}{\nonumber}
\newcommand {\pt}{\partial}

\title{\boldmath Localization of matter fields on a chameleon brane}

\author[a,b]{Yi Zhong,}
\author[c,1]{and Ke Yang\note{Corresponding author.}}

\affiliation[a]{ School of Physics and Electronics Science,\\
Hunan Provincial Key Laboratory of High-Energy Scale Physics and Applications,\\
             Hunan University, Changsha 410082, P. R. China}
\affiliation[b]{Lanzhou Center for Theoretical Physics,
Key Laboratory of Theoretical Physics of Gansu Province,
School of Physical Science and Technology,
Lanzhou University, Lanzhou 730000, P. R. China}
\affiliation[c]{School of Physical Science and Technology, Southwest University, Chongqing 400715, P. R. China}

\emailAdd{zhongy@hnu.edu.cn}
\emailAdd{keyang@swu.edu.cn}

\abstract{In this work, we address the localization problem of vector field in the chameleon braneworld and investigate the localization of various matter fields. The conditions for localizing the matter fields are determined. It is found that the zero modes of scalar, vector, and fermion fields can be successfully localized, yet the zero mode of Kalb-Ramond field cannot be localized, which implies that the recovery of standard model fields on the brane. Furthermore, the characteristics of quasi-localized modes of the $q$-form fields are analyzed, and the parameter constraints of the model are estimated.}

\begin{document}
\maketitle
\flushbottom

\section{Introduction}
The existence of hidden extra dimensions in our spacetime, particularly at ultraviolet scales, has been a topic of interest for a long time, starting from the proposal of the Kaluza-Klein (KK) theory in the 1920s. Based on this idea, the braneworld scenario was proposed, in which our universe could be a 3-brane embedded in a higher-dimensional space-time, called the bulk.
This perspective offers a novel approach to addressing the gauge hierarchy problem and cosmological constant problem \cite{ArkaniHamed:1998rs,ArkaniHamed:2000eg,Randall:1999ee,Randall:1999vf}, which are two persistent challenges in particle physics and cosmology. The concept of a thick brane \cite{Gremm:2000dj,Gass:1999gk,Gremm:1999pj,Afonso:2006gi} extends the ideas of the Randall-Sundrum-2 (RS2) model \cite{Randall:1999vf}, and the configuration of a domain wall \cite{Rubakov:1983bb,Rubakov:1983bz} serves as a topological defect in the bulk.

In the braneworld theory, the localization of matter fields is an important issue, as it not only allows the five-dimensional theory to reduce to a four-dimensional effective theory at low energy scales but also provides observable effects for probing the extra dimensions \cite{Bajc2000,Oda2000,Melfo:2006hh,Liu:2009ve,PhysRevD.79.125022,Liang2009,German:2013sk,Vaquera-Araujo:2014tia,Salvio:2007qx}. In five-dimensional spacetime, free scalar fields can typically be localized on branes \cite{Bajc2000,Oda2000}, while fermion fields generally require Yukawa couplings with background scalar fields to achieve localization on branes \cite{Melfo:2006hh,PhysRevD.78.065025,Liang2009,Alencar2014,Vaquera-Araujo:2014tia}. However, it is generally challenging to localize vector field on branes in five dimensions. Some researchers have attempted to address this issue by assuming couplings between vector field and scalar field or spacetime geometry, or by considering models such as the Weyl geometry braneworld, de Sitter braneworld, and six-dimensional braneworld  \cite{Chumbes:2011zt,Alencar2014,Vaquera-Araujo:2014tia}.
See Refs.~\cite{Dvali:2000rx,Akhmedov:2001ny,Guerrero:2009ac,Delsate:2011aa,Cruz:2012kd,Dimopoulos:2000ej,Belchior:2023gmr,Zhao:2017epp} for more works on the localization of vector field in braneworld models. Despite the extensive research conducted on this issue, there is still significance in providing a concise and natural new mechanism to effectively localize the vector field on the brane.

On the other hand, the chameleon gravity proposed by Khoury et al. in 2003 can address the dark energy problem \cite{PhysRevLett.93.171104,PhysRevD.69.044026}. This theory introduces a scalar field, known as the "chameleon" which exhibits mass variation depending on the surrounding matter density. The concept of chameleon gravity is closely related to string theory \cite{Brax2004} and scalar-tensor theory. The theory has been extensively tested through observations and experiments \cite{Burrage:2017qrf}, showing promising prospects for resolving the Hubble constant problem \cite{PhysRevD.103.L121302}. In this theory, matter fields do not directly couple to the metric tensor $g_{\mu\nu}$ that describes spacetime curvature. Instead, they couple to a physical metric $\widetilde{g}_{\mu\nu}=b^2(\phi)g_{\mu\nu}$ resulting from a conformal transformation with the chameleon scalar field $\phi$. Thus, various matter fields are coupled to the chameleon scalar field $\phi$, providing a potential solution to the localization problem of vector field. Moreover, the coupling of scalar fields and fermion fields in matter to the chameleon scalar field $\phi$ results in novel characteristics within the braneworld model in chameleon gravity. Brane cosmology in chameleon gravity was investigated in Refs.~\cite{PhysRevD.85.023526,Bisabr2017}, where thin brane models were considered. While thick brane model in chameleon gravity has not yet been explored, this study aims to address this gap. We will construct a thick chameleon brane and attempt to solve the localization problem of vector field in the chameleon thick braneworld. Furthermore, we will investigate the localization property of various matter fields  in this model.

The layout of the paper is as follows: In Section 2, we construct a  chameleon brane model within the background of a Sine-Gordon kink. In Section 3, we explore the localization and quasi-localization of three types of $q$-form fields. In Section 4, we examine the localization of a fermion field. Finally, we conclude with a discussion of our findings. Throughout this paper, $M,N,\cdots $ denote the indices of the five dimensional coordinates, and $\mu,\nu,\cdots$ denote the ones on the brane.

\section{Braneworld in chameleon gravity}
We start with the following five-dimensional action
\begin{equation}
    S=\int d^5 x \sqrt{-g}\left[ \frac{1}{2}R-\frac{1}{2}\nabla_M \phi \nabla^M \phi -V(\phi) \right]+ S_m [\tilde{g}_{MN},\psi_i],
\end{equation}
which describes a canonical scalar field minimally coupled to gravity, and  matter fields $\psi_i$ coupled to the Jordan frame metric
$\tilde{g}_{MN}=b^2 (\phi) g_{MN}$.   It can be seen from the above action that the braneworld solution in this set up is the same with the standard case. However, there are  some novel features regarding the localization of various matter fields are expected, since the matter fields couple to the Jordan frame metric $\tilde{g}_{MN}$ instead of the Einstein frame one $g_{MN}$.

The metric ansatz of a flat brane embedded in a five-dimensional anti-de Sitter (AdS) spacetime is
\begin{equation}
\label{metric}
g_{MN}=a^2 (z)
\left[ \eta_{\mu\nu}dx^{\mu}dx^{\nu}+dz^2
\right],
\end{equation}
where $a(z)=e^{A(z)}$ is the warped factor.
Varying the action with respect to the Einstein frame metric $g_{MN}$ and the scalar field $\phi$, and let $S_m=0$, we obtain the following equation of motion,
\begin{eqnarray}
3A'^2-3A'' &=& \phi'^2, \\
12A'^2+2\text{e}^{2A}V(\phi) &=& \phi'^2,    \\
\phi''+3A'\phi' &=& \text{e}^{2A}\frac{\pt V(\phi)}{\pt \phi}.
\end{eqnarray}
Assuming the sine-Gordon potential $V(\phi)=\frac{3}{4} k^2 \left[5 \cos \left(\frac{2 \phi }{\sqrt{3}}\right)-3\right]$, the brane solution is solved as \cite{Koley:2004at}
\begin{eqnarray}
A(z) &=& -\frac{1}{2} \ln \left(k^2 z^2+1\right),    \\
\phi(z) &=& \sqrt{3}\arctan \left(kz\right),
\end{eqnarray}
where the parameter $k$ is associated with the inverse of the  five-dimensional  AdS radius.
Since the brane world solution is the same with the standard case, the KK gravitons generate a correction, which is proportional to ${1}/{kr^3}$, to the  Newtonian potential \cite{Csaki2000,Bazeia2009,Liu:2012rc,Yang:2022fed}.
Conversely, a recent experiment has demonstrated that the length scale deviating from the gravitational inverse square law is at most $48 \mu \text{m}$ \cite{PhysRevLett.124.051301}.
Thus the parameter $k$ can be roughly estimated to be $k\gtrsim 10^{-3}\text{eV}$.

In the thick brane models, the extra dimension is non-compact, and the matter fields are localized on the brane by the warped geometry, which is described by the Jordan frame metric
\begin{equation}
\label{tilde g}
\tilde{g}_{MN}=\tilde{a}^2(z)
\left[ \eta_{\mu\nu}dx^{\mu}dx^{\nu}+dz^2
\right],
\end{equation}
where we have defined an effective warped factor $\tilde{a}(z)=\text{e}^{\tilde{A}(z)}=\text{e}^{B(\phi)+A(z)}$ and $b(\phi)=\text{e}^{B(\phi)}$ for convenience, and all the functions deduced from $\tilde {g}_{MN}$ are with tilde in the following.

\section{Localization of $q$-form fields}
In this section, we consider the localization of  $q$-form field $X_{M_1 M_2 \ldots M_q}$ on the brane, which represents a scalar field for $q=0$, a vector field for $q=1$, and Kalb-Ramond (KR) field for $q=2$. We will first investigate a generic $q$-form field, and then discuss each of the scalar, vector and Kalb-Ramond fields.
The action of a massless $q$-form field in the bulk is given by
\begin{equation}
\label{q-form action}
    S_q = -\frac{1}{2(q+1)!} \int d^5 x \sqrt{-\tilde{g}} Y^{M_1 M_2 \ldots M_{q+1}} Y_{M_1 M_2 \ldots M_{q+1}},
\end{equation}
where $\tilde{g}$ is the determinant of the Jordan frame metric $\tilde{g}_{MN}$, the indices are also raised and lowered by $\tilde{g}_{MN}$, and the strength of the $q$-form  field  is defined as $Y_{M_1 M_2 \ldots M_{q+1}}=\pt_{[M_1} X_{M_2 M_3 \ldots M_{q+1}]}$.
By varying action (\ref{q-form action}) with respect to $X_{M_1 M_2 \ldots M_q}$, the equations of motion of the $q$-form field can be obtained,
\begin{eqnarray}
\label{q-form eom1}
       \pt_{\mu_1} \left(\sqrt{-\tilde{g}} Y^{\mu_1 \mu_2 \ldots \mu_{q+1}} \right) + \pt_{z} \left(\sqrt{-\tilde{g}} Y^{z \mu_2 \ldots \mu_{q+1}} \right) &=&0,\\
    \label{q-form eom2}
        \pt_{\mu_1} \left(\sqrt{-\tilde{g}} Y^{\mu_1 \ldots \mu_{q-1} z} \right)&=&0.
\end{eqnarray}
\subsection{Localization conditions of the zero modes}
Before we employing the Kaluza-Klein (KK) decomposition, it is essential to eliminate the gauge degrees of freedom of the $q$-form field. It is obvious that the action (\ref{q-form action}) is invariant under the gauge transformation $X_{M_1 M_2 \ldots M_q} \rightarrow \tilde{X}_{M_1 M_2 \ldots M_q}=X_{M_1 M_2 \ldots M_q}+\pt_{[M_1} \Lambda_{M_2 M_3 \ldots M_{q}]}$  with $\Lambda_{M_2 M_3 \ldots M_{q}}$ an antisymmetric tensor field. Therefore, one can eliminate the gauge degrees of freedom by assuming $X_{\mu_1 \ldots \mu_{q-1} z}=0$. Then we employ the KK decomposition for the rest components,
\begin{eqnarray}
\label{q-form kk}
    X_{\mu_1 \mu_2\ldots \mu_{q}}(x, z)= \sum_{n} \hat{X}_{\mu_1 \mu_2\ldots \mu_{q}}^{(n)}(x)\Upsilon_n(z)\text{e}^{\frac{(2q-3)\tilde{A}}{2}}.
\end{eqnarray}	
The corresponding KK decomposition for the strength is
\begin{eqnarray}
\label{q-form kk2}
    Y_{\mu_1 \mu_2 \ldots \mu_{q+1}}(x, z) &=& \sum_{n} \hat{Y}_{\mu_1 \mu_2 \ldots \mu_{q+1}}^{(n)}(x)\Upsilon_n(z) \text{e}^{\frac{(2q-3)\tilde{A}}{2}},   \nn\\
    Y_{\mu_1 \mu_2 \ldots \mu_q z}(x, z)&=&\sum_{n} \hat{Y}_{\mu_1 \mu_2\ldots \mu_{q}}^{(n)}(x) \Upsilon_n(z) \text{e}^{\frac{(2q-3)\tilde{A}}{2}}\left(\Upsilon'_{n}(z)+\frac{(2q-3)}{2} \tilde{A}'\Upsilon_{n}(z) \right).
\end{eqnarray}	
Substituting the above decompositions into the Eqs.~(\ref{q-form eom1}) and (\ref{q-form eom2}), it is easy to obtain the Schr\"odinger-like equation of the  KK modes $\Upsilon_n(z)$,
 \begin{eqnarray}
\label{q-form schro}
    \left[ -\pt^2  _z +V_q (z) \right] \Upsilon_n(z)=m_n ^2 \Upsilon_n(z),
\end{eqnarray}	
 with the effective potential $V_q (z)$ given by
  \begin{eqnarray}
\label{q-form potential}
    V_q (z) = \frac{3-2q}{2}\tilde{A}'' + \frac{(3-2q)^2}{4} \tilde{A}'^2.
\end{eqnarray}	
Therefore, we can solve the zero mode with $m_0 ^2=0$, yielding
  \begin{eqnarray}
\label{vector 0}
 \Upsilon_{0}(z)=N_0 \text{e}^{\frac{(3-2q)\tilde{A}}{2}},
\end{eqnarray}	
where $N_0$ is a normalization constant.
In order to ensure that the five-dimensional action (\ref{q-form action}) reduces to a four-dimensional effective action on the brane, the five-dimensional $q$-form field has to be localized on the brane,  which can be realized by introducing the orthonormal condition
\begin{eqnarray}
\label{orth condition}
    \int^{+\infty}_{-\infty} \Upsilon_{m}(z) \Upsilon_{n}(z) dz = \delta_{mn}.
\end{eqnarray}
Then the action (\ref{q-form action}) can be reduced to the following effective action:
\begin{eqnarray}
\label{q-form eff action}
 \!\!\!\!\!\!\!S_q\!=\!-\frac{1}{2(q+1)!} \sum_{n}\! \int \!d^4 x \!\left(\hat{Y}_{\mu_1 \mu_2 \ldots \mu_{q+1}}^{(n)} \hat{Y}^{(n) \mu_1 \mu_2 \ldots \mu_{q+1}} \!+\!\frac{M_n^2}{q+1}\hat{X}_{\mu_1 \mu_2 \ldots \mu_{q}}^{(n)} \hat{X}^{(n) \mu_1 \mu_2 \ldots \mu_{q}} \!\right)\!,
\end{eqnarray}
where the indices are raised and lowered by the Minkowski metric $\eta_{\mu\nu}$.  The  effective action describes a massless $q$-form field ($n=0$) and a series of massive $q$-form fields  ($n\geq 1$) .
Therefore,  the condition that the zero mode is localized on the brane is given by
\begin{eqnarray}
    \int^{+\infty}_{-\infty} \Upsilon_0^2 (z) dz
    = N_0^2 \int^{+\infty}_{-\infty} \left[a(z)b(\phi)\right]^{3-2q} dz =1,
\end{eqnarray}
or,
\begin{eqnarray}
\label{q-form condition1}
 	\int^{+\infty}_{-\infty} \left[a(z)b(\phi)\right]^{3-2q} dz <\infty.
\end{eqnarray}
The above condition can be rewritten as $\left[a(z)b(\phi)\right]^{3-2q} \rightarrow z^p$ in which $p<-1$.
 Furthermore, note that $a(\pm \infty)=\pm\frac{1}{ kz}$, so the localization condition reduces to
\begin{eqnarray}
\label{q-form condition2}
  b[\phi(\infty)] = z^{\frac{p}{3-2q}+1}
\end{eqnarray}
where $p<-1$. In addition, to prevent that Eq.~(\ref{q-form condition1}) vanishes and for the consideration of symmetry, $b(\phi)$ must be an even function of $\phi$.

Next, we focus on the cases of the scalar field ($q=0$),  vector field ($q=1$), and KR field ($q=2$), respectively.

\subsubsection{Scalar field}
For the case $q=0$, the action (\ref{q-form action}) represents a massless scalar field $\Phi(x^{\mu},z)$, of which the KK composition is
\begin{eqnarray}
\label{scalar kk}
    \Phi(x^{\mu},z) = \sum_n \varphi^{(n)}(x^{\mu})\chi_{n}(z) \text{e}^{-\frac{3}{2}\tilde{A}}.
\end{eqnarray}
The effective potential reads
\begin{eqnarray}
\label{scalar potential}
    V_{\text{s}} (z) = \frac{3}{2}\tilde{A}'' + \frac{9}{4} \tilde{A}'^2.
\end{eqnarray}
 If the orthonormal condition $\int dz \chi_{m} \chi_{n} =\delta_{nm}$ is satisfied, the fundamental five-dimensional action of a free massless scalar filed can be reduced to the four-dimensional action of a massless and a series of massive scalar fields, given by
\begin{eqnarray}
\label{scalar eff action}
    S_{\text{s}}=-\sum_{n\geq1} \int d^4 x \left(\frac{1}{2}\pt^{\mu} \varphi^{(0)} \pt_{\mu}\varphi^{(0)}+\frac{1}{2}\pt^{\mu} \varphi^{(n)} \pt_{\mu}\varphi^{(n)}+m_{(n)}^2 \varphi^{(n)}\varphi^{(n)}  \right).
\end{eqnarray}
In order to be consistent with the standard model, localization of the zero mode $\chi_{0} (z)=\text{e}^{\tilde{A}}$ is essential. From Eq.~(\ref{q-form condition2}) we can obtain the localization condition is
\begin{eqnarray}
\label{scalar condition2}
  b[\phi(\infty)] = z^{r},
\end{eqnarray}
with $r<\frac{2}{3}$.
\subsubsection{Vector field}
Then we consider the case $q=1$, the action (\ref{q-form action}) represents a U(1) gauge vector field.
We choose the gauge $A_5=0$, and make the KK decomposition for the vector field
\begin{eqnarray}
\label{vector kk}
    A_{\mu}(x, z)= \sum_{n} a^{(n)}_{\mu}(x)\alpha_{n} (z) \text{e}^{-\frac{\tilde{A}}{2}}.
\end{eqnarray}	
 With the above decomposition, it is easy to obtain the Schr\"odinger-like equation of the Vector KK modes $\alpha_{(n)} (z)$:
 \begin{eqnarray}
\label{vector schro}
    \left[ -\pt^2  _z +V_{\text{v}} (z) \right] \alpha_{n} (z)=m_n ^2 \alpha_{n} (z),
\end{eqnarray}	
 with the effective potential $V_{\text{v}} (z)$ given by
  \begin{eqnarray}
\label{vector potential}
    V_{\text{v}} (z) = \frac{1}{2}\tilde{A}'' + \frac{1}{4} \tilde{A}'^2.
\end{eqnarray}	
Therefore, the massless vector zero mode reads
  \begin{eqnarray}
\label{vector 0}
    \alpha_{0} (z)=N_0 \text{e}^{\frac{\tilde{A}}{2}}.
\end{eqnarray}	
where $N_0$ is a normalization constant. Now the orthonormal condition (\ref{orth condition}) is written as
\begin{eqnarray}
\label{vector condition}
    \int^{+\infty}_{-\infty} \alpha_{m} (z) \alpha_{n} (z) dz = \delta_{mn}.
\end{eqnarray}
Then the action (\ref{q-form action}) can be reduced to the following effective action:
\begin{eqnarray}
\label{vector eff action}
  S_\text{v}=-\sum_{n\geq1}\int d^4 x \left( \frac{1}{4} f^{(0)\mu\nu}f^{(0)}_{\mu\nu} +\frac{1}{4}f^{(n)\mu\nu}f^{(n)}_{\mu\nu}+\frac{1}{2}m_n^2a^{(n)\mu} a^{(n)}_{\mu} \right),
\end{eqnarray}
which describes a massless vector field and a series of massive vector fields.

From Eq.~(\ref{q-form condition2})  the localization of the U(1) gauge field on the brane requires
\begin{eqnarray}
\label{vector condition}
  b[\phi(\infty)] = z^{r},
\end{eqnarray}
with $r<0$.

\subsubsection{KR field}
At last, we consider the case $q=2$,  and the action (\ref{q-form action}) represents a massless KR field $B_{MN}$. The KK decomposition (\ref{q-form kk}) reduces to
\begin{eqnarray}
\label{KR kk}
    B_{\mu\nu}(x, z)= \sum_{n} b^{(n)}_{\mu\nu}(x)u_n (z) \text{e}^{\frac{\tilde{A}}{2}},
\end{eqnarray}
and the effective potential (\ref{q-form potential}) reads
\begin{eqnarray}
\label{KR potential}
   V_{\text{KR}}(z)=\frac{1}{4}A'^2-\frac{1}{2}A''.
\end{eqnarray}
If the orthonormal condition (\ref{q-form condition2}) is satisfied, the action of the five-dimensional KR field reduces to
\begin{eqnarray}
\label{vector eff action}
    S_{\text{KR}}= -\sum_{n\geq1} \int d^4 x \left( F^{(0) \mu\nu\lambda}F^{(0)}_{\mu\nu\lambda}+F^{(n) \mu\nu\lambda}F^{(n)}_{\mu\nu\lambda}+\frac{1}{3}m_n^2 b^{(n) \mu\nu}  b^{(n)}_{\mu\nu} \right),
\end{eqnarray}
which describes a four-dimensional massless KR field and a series of four-dimensional massive KR fields on the brane.

From Eq.~(\ref{q-form condition2}), we obtain the localization condition of the zero mode, i.e.,
\begin{eqnarray}
\label{vector condition}
  b[\phi(\infty)] = z^{r},
\end{eqnarray}
with $r>0$.

In order to be consistent with the observations that photons are the carrier of the electromagnetic interaction but KR particles have not been discovered yet on the brane, it is necessary for the vector field to be localized but KR filed to not be localized on the brane. Interestingly, the condition for localizing the vector field conflicts with that of the KR field, and therefore the expected result can indeed be achieved as long as $r<0$.

In summary, in the chameleon braneworld, the scalar and vector fields can be localized on the brane, while the KR field can not. The localization condition of scalar and vector fields is:
\begin{equation}
\label{q-form condition3}
 	\int^{+\infty}_{-\infty} a(z) b(\phi) dz <\infty.
\end{equation}
It is clear that $b(\phi)$ can not be an odd function of $\phi$, otherwise the integration vanishes. For simplicity, here we assume that $b(\phi)$ is an even function of $\phi$. As an example, the above condition can be simply satisfied by just choosing $b(\phi)=\cos^p(\frac{\phi}{\sqrt{3}})$ with $p>0$, which is the ansatz we adopted in the rest of the work.

\subsection{Quasi-localiztion of $q$-form fields}
In addition to the above investigation of the bounded zero modes, it is also necessary to consider the mass spectra of the quasi-localized states (i.e., resonances).  Since the effective potential (\ref{q-form potential}) is an even function of $z$, the resonance $\Upsilon_n (z)$ has either an even parity $\Upsilon^{(n)}_{\text{e}} (z)$ or an odd parity $\Upsilon^{(n)}_{\text{o}} (z)$.  They respectively satisfy the following boundary conditions,
    \begin{eqnarray}
        \label{condition even}
        \Upsilon^{(n)}_{\text{e}} (0)=1,~~~~~~~~\partial_z \Upsilon^{(n)}_{\text{e}}  (0)=0;\\
        \label{condition odd}
        \Upsilon^{(n)}_{\text{o}} (0)=0,~~~~~~~~\partial_z \Upsilon^{(n)}_{\text{o}} (0)=1.
    \end{eqnarray}
The resonances can be found by adopting the concept of the relative probability of the KK mode $\Upsilon^{(n)}(z)$ with mass $m_n$, which was defined in Ref.~\cite{Liu:2009ve,PhysRevD.79.125022}:
    \begin{eqnarray}
        \label{relative p1}
        P(m_n^2)=\frac{\int^{z_b}_{-z_b}|\Upsilon^{(n)}(z)|^2 dz}{\int^{z_{max}}_{-z_{max}}|\Upsilon^{(n)}(z)|^2 dz}.
    \end{eqnarray}
Here $2z_b$ is approximately the width of the thick brane (or the width of the effective potential well $V_q (z)$), and $z_{max}=10z_b$.
By solving the Schr\"{o}dinger-like equation (\ref{q-form schro}) for a given $m_n^2$ numerically, the corresponding relative probability $P$ can be obtained for the even or odd modes. Then each peak in the figure of $P(m_n^2)$ represents a resonant mode.  The life-time $\tau$ of
 a resonant mode can obtained by $\tau=\frac{1}{\Gamma}$, where $\Gamma$ is the full width at half
 maximum (FWHM) \cite{PhysRevLett.84.5928,PhysRevD.79.125022}.
It is more convenient to use the dimensionless quantities $m/k$ and $k\tau$, which have been scaled with $k$.
 Next we analyze the resonances of scalar, vector and KR fields, respectively.

   \begin{figure}[h]
    \begin{center}
    \subfigure[Scalar field]{\label{VsScalar}
        \includegraphics[width=4.83cm]{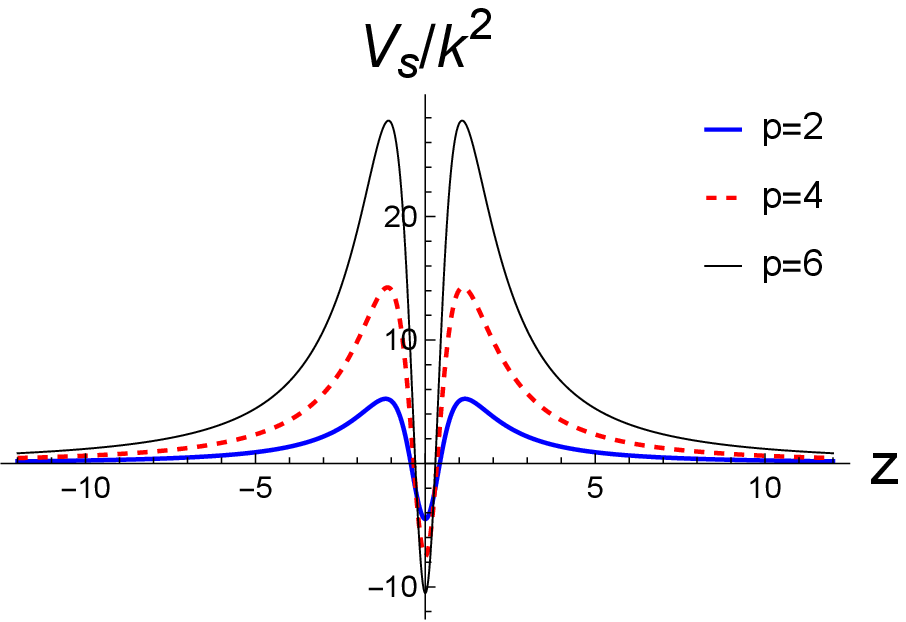}}
    \subfigure[Vector field]{\label{Vvector}
        \includegraphics[width=4.83cm]{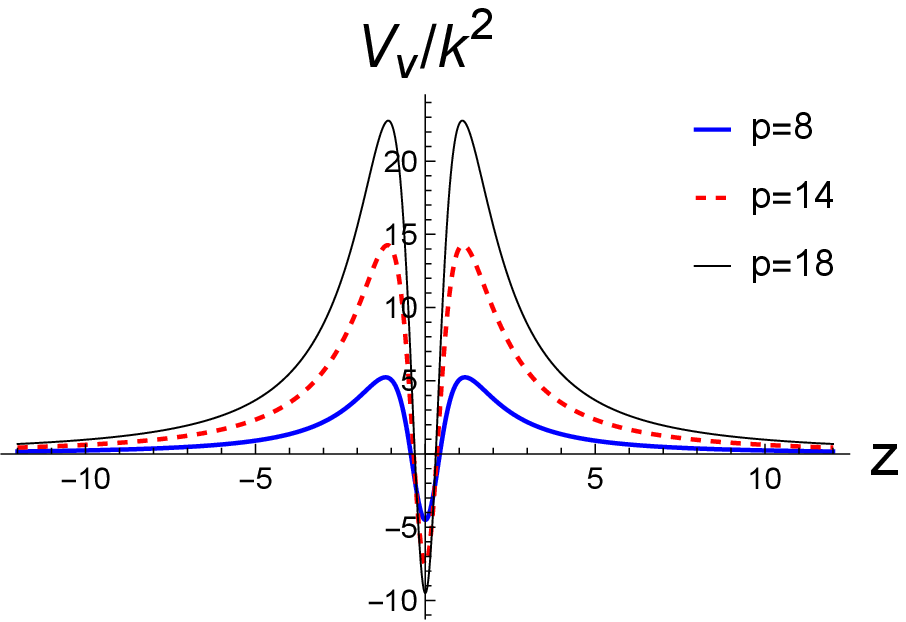}}
    \subfigure[KR field]{\label{VKR}
        \includegraphics[width=4.83cm]{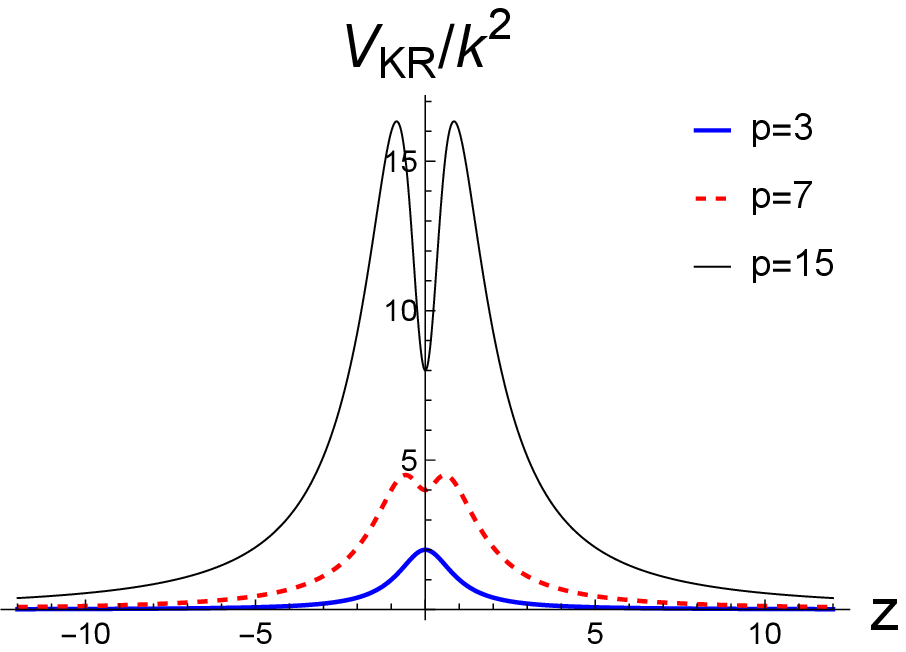}}
    \end{center}
    \caption{The effective potential  of the q-form fields for different values of $p$. }  \label{Pscalar}
    \end{figure}

\subsubsection{Scalar field}
\begin{figure}[h]
    \begin{center}
    \subfigure[relative probability $P_{max}$]{\label{figure P 31 }
        \includegraphics[width=4.83cm]{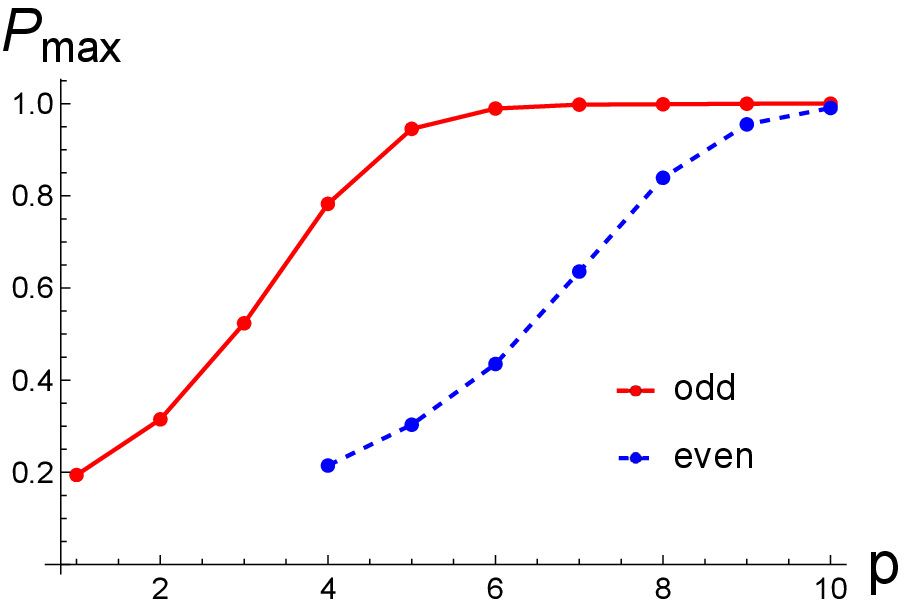}}
    \subfigure[scaled mass ${m}/{k}$]{\label{figure P 32}
        \includegraphics[width=4.83cm]{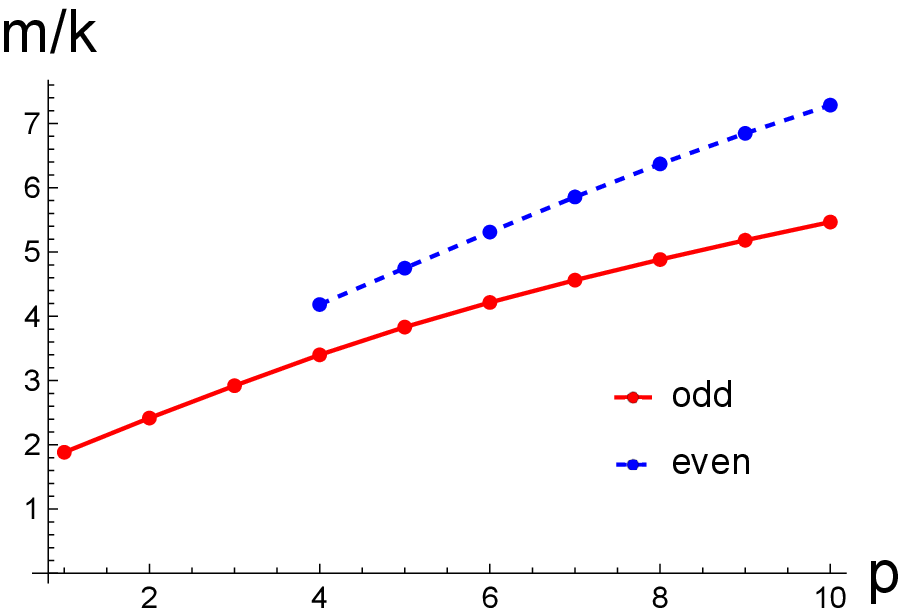}}
    \subfigure[scaled life-time $k\tau$]{\label{figure P 33}
        \includegraphics[width=4.83cm]{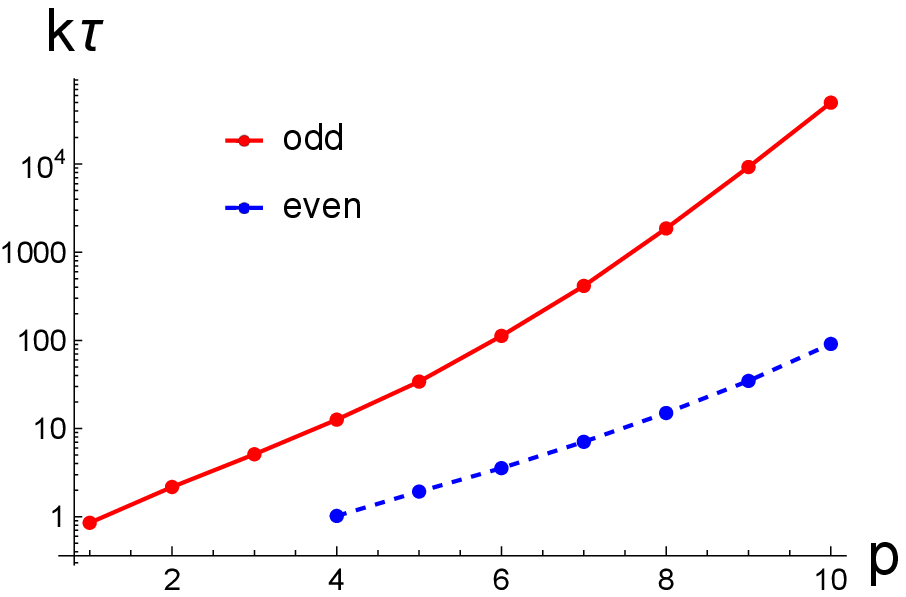}}
    \end{center}
    \caption{The relative probability $P_{max}$, scaled mass ${m}/{k}$,  and scaled life-time $k\tau$  of the resonances of scalar field for different values of $p$.}  \label{PMLScalar}
    \end{figure}

The effective potential of the scalar field  and the relative probability as a function of $m^2/k^2$ are  plotted in Fig.~\ref{VsScalar} and Fig.~\ref{Pscalar} for different values of $p$, respectively. Since the maximum of the potential increases with the parameter $p$, it can be expected that the relative probability, mass, and life-time of the resonances also increases with the parameter $p$, as is shown in Fig.~\ref{PMLScalar}. As shown in the figures, the first scalar resonance, which has an old parity, appears only if $p\gtrsim 1$. Since the bound zero mode has an even parity, the first resonance is an odd mode.

   \begin{figure}[h]
    \begin{center}
    \subfigure[$p=2$]{\label{figure P 31 }
        \includegraphics[width=4.83cm]{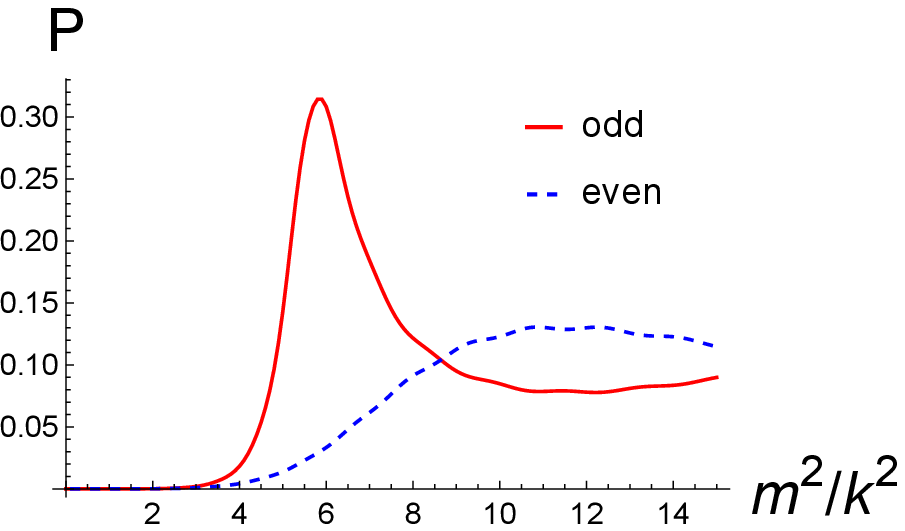}}
    \subfigure[$p=4$]{\label{figure P 32}
        \includegraphics[width=4.83cm]{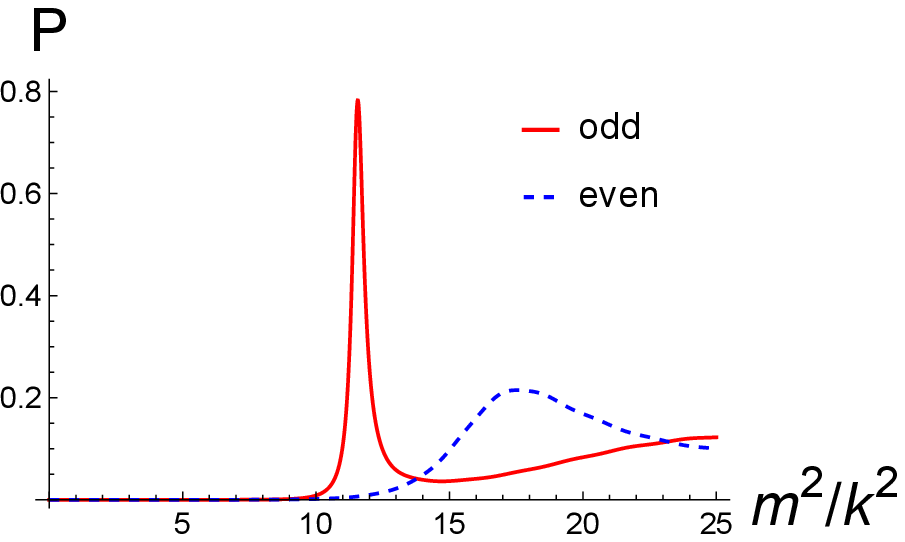}}
    \subfigure[$p=6$]{\label{figure P 33}
        \includegraphics[width=4.83cm]{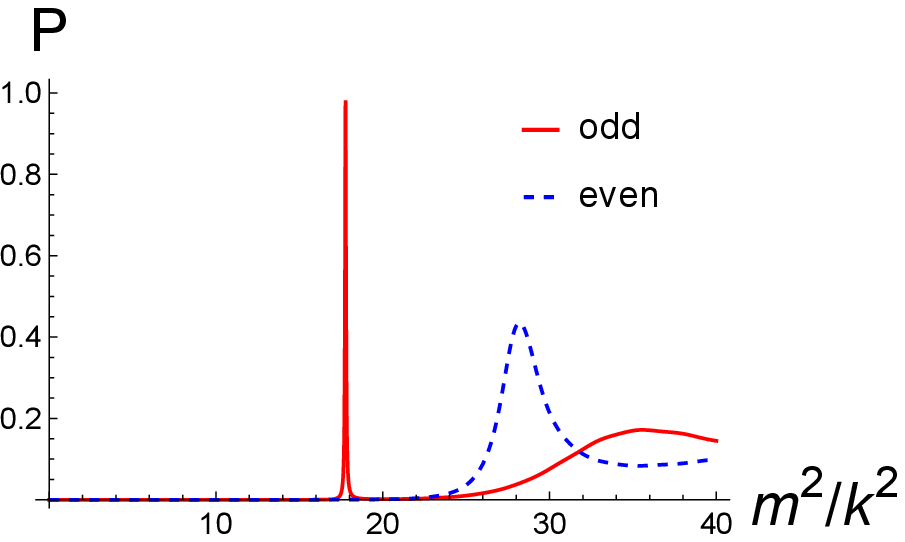}}
    \end{center}
    \caption{The relative  probability  $P\left({m^2}/{k^2}\right)$ of the even and odd parity modes of scalar field for different values of $p$. }  \label{Pscalar}
    \end{figure}

\subsubsection{Vector field}
\begin{figure}[htb]
    \begin{center}
    \subfigure[$p=8$]{\label{figure P 31 }
        \includegraphics[width=4.83cm]{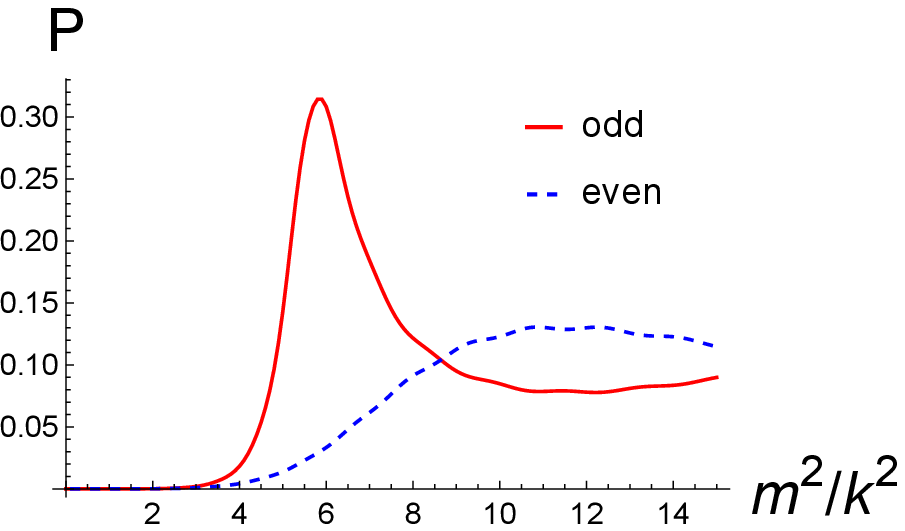}}
    \subfigure[$p=14$]{\label{figure P 32}
        \includegraphics[width=4.83cm]{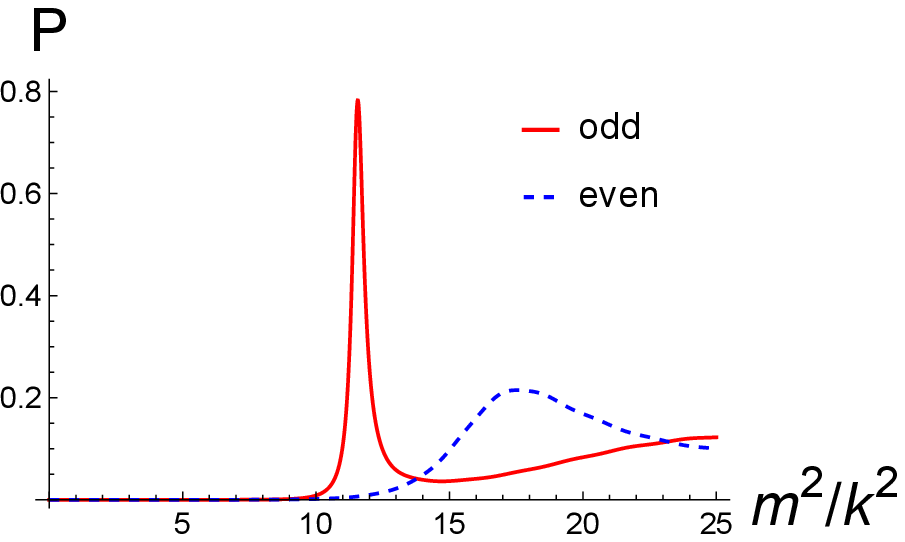}}
    \subfigure[$p=18$]{\label{figure P 33}
        \includegraphics[width=4.83cm]{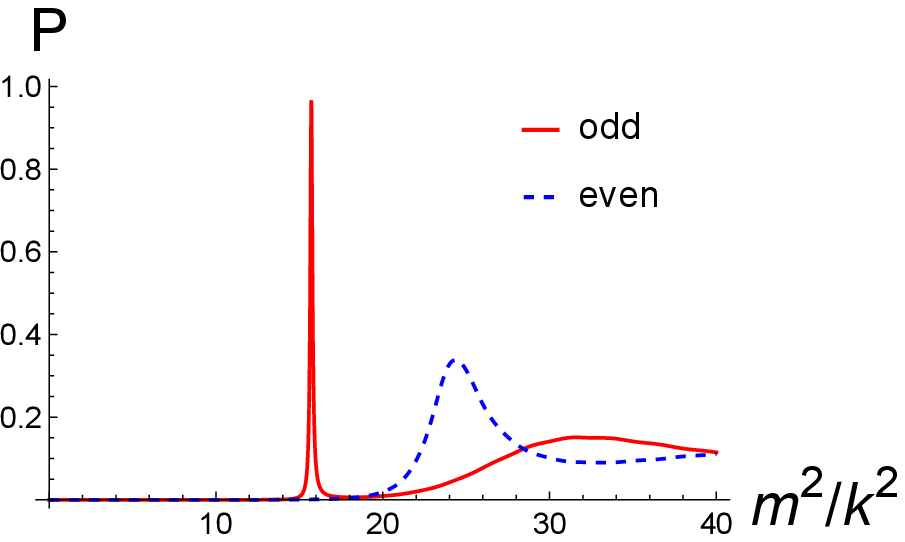}}
    \end{center}
    \caption{The relative  probability  $P\left({m^2}/{k^2}\right)$ of the even and odd parity modes of the vector  field for different values of $p$. }  \label{Pvector}
    \end{figure}

\begin{figure}[htb]
    \begin{center}
    \subfigure[relative probability $P_{max}$]{\label{figure P 31 }
        \includegraphics[width=4.83cm]{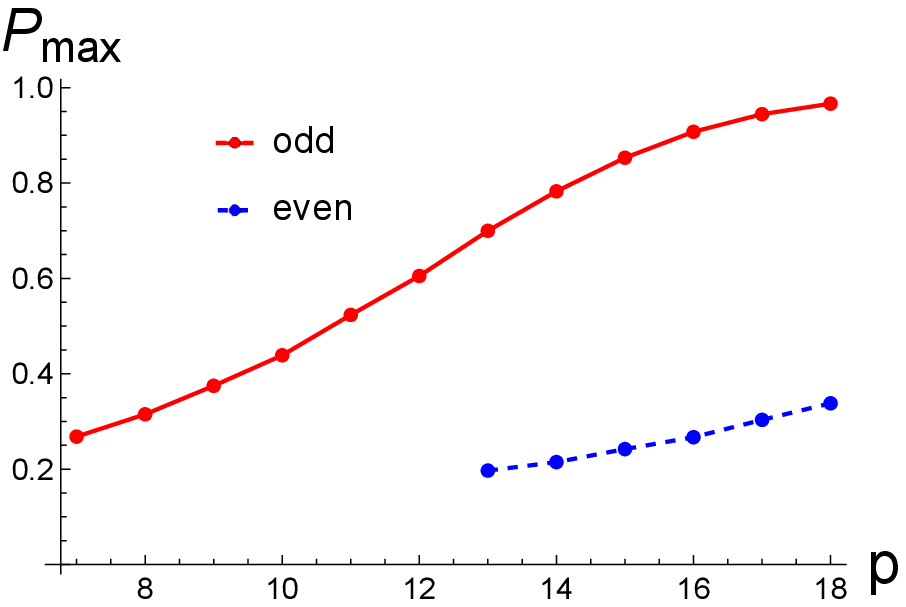}}
    \subfigure[scaled mass ${m}/{k}$]{\label{figure P 32}
        \includegraphics[width=4.83cm]{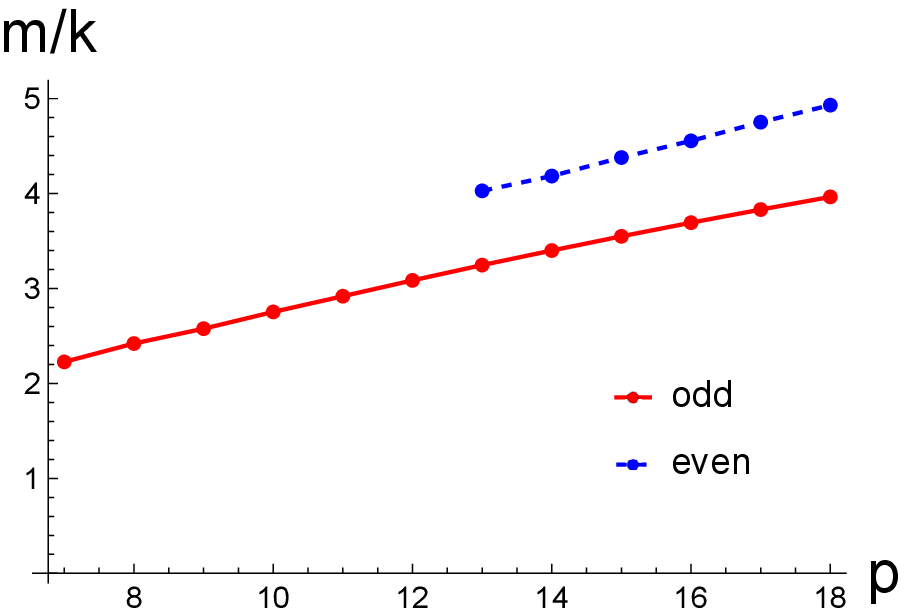}}
    \subfigure[scaled life-time $k\tau$]{\label{figure P 33}
        \includegraphics[width=4.83cm]{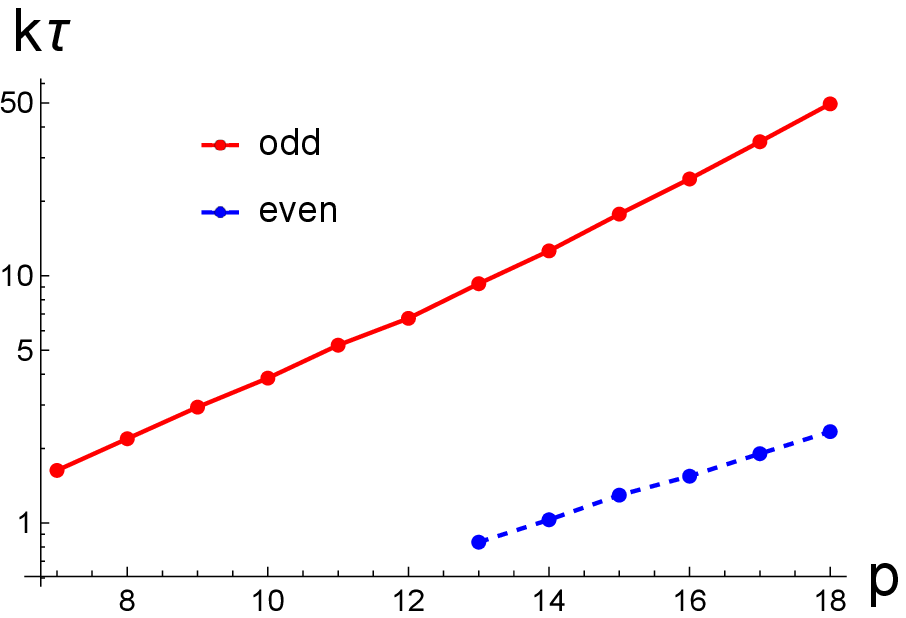}}
    \end{center}
    \caption{The relative probability $P_{max}$, scaled mass ${m}/{k}$,  and scaled life-time $k\tau$  of the resonances of the vector  field for different values of $p$.}  \label{PMLVector}
    \end{figure}

The resonances of the vector field behave similar with those of the scalar field. The effective potential of the vector field  and the relative probability as a function of $m^2/k^2$ are  plotted in Fig.~\ref{Vvector} and Fig.~\ref{Pvector} for different values of $p$, respectively. And the relative probability, mass, and life-time of the resonances for different values of the parameter $p$ is shown in Fig.~\ref{PMLVector}. As shown in the figures, the first vector resonance, which processes an old parity, appears when $p\gtrsim 7$.

\subsubsection{KR field}
\begin{figure}[htb]
    \begin{center}
    \subfigure[$p=3$]{\label{figure P 31 }
        \includegraphics[width=4.83cm]{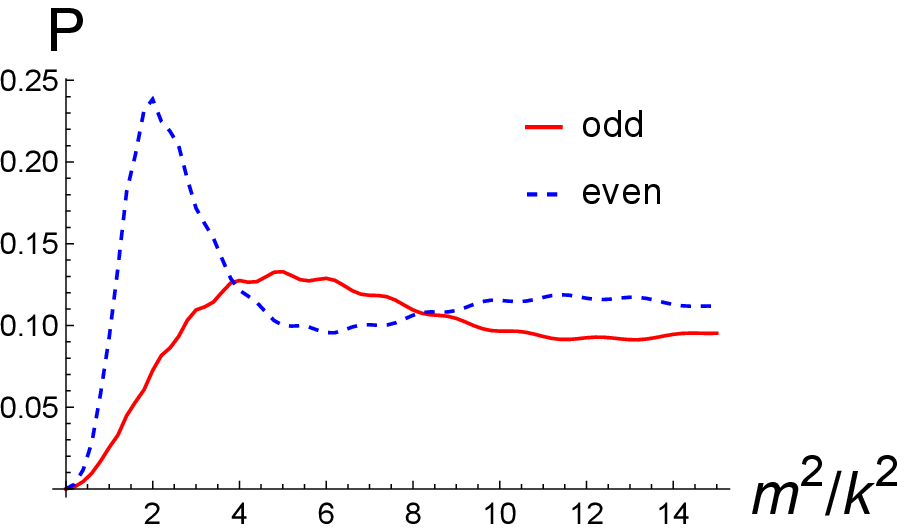}}
    \subfigure[$p=7$]{\label{figure P 32}
        \includegraphics[width=4.83cm]{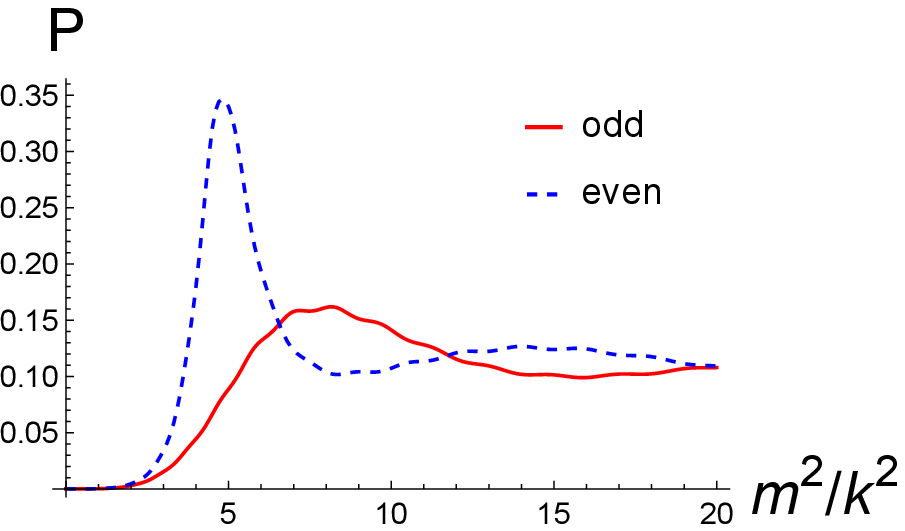}}
    \subfigure[$p=15$]{\label{figure P 33}
        \includegraphics[width=4.83cm]{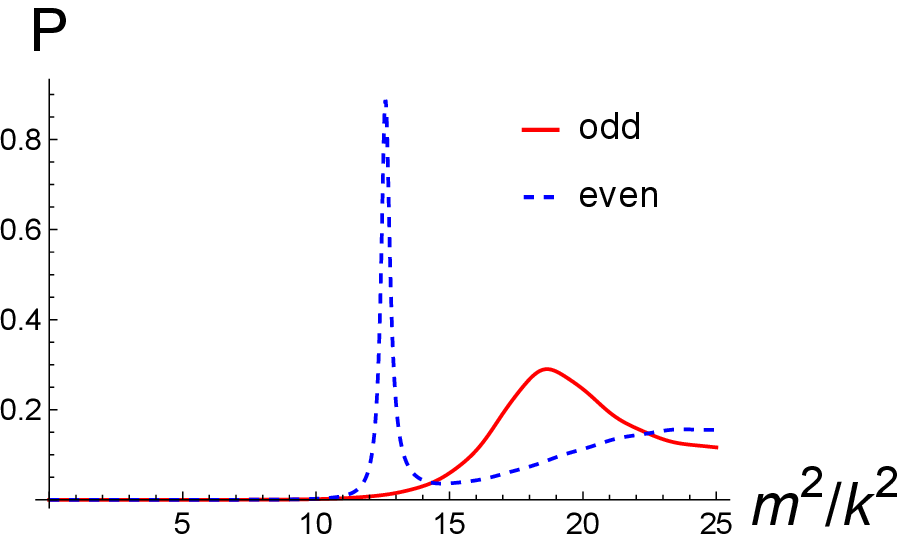}}
    \end{center}
    \caption{The relative  probability  $P\left(m^2/k^2\right)$ of the even and odd parity modes of the KR  field for different values of $p$. }  \label{PKR}
    \end{figure}

\begin{figure}[htb]
    \begin{center}
    \subfigure[relative probability $P_{max}$]{\label{figure P 31 }
        \includegraphics[width=4.83cm]{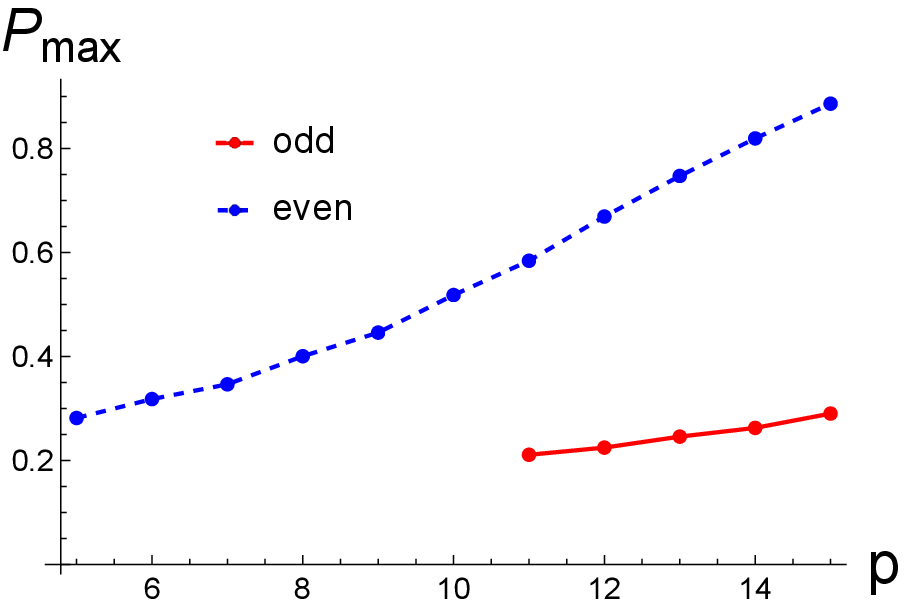}}
    \subfigure[scaled mass ${m}/{k}$]{\label{figure P 32}
        \includegraphics[width=4.83cm]{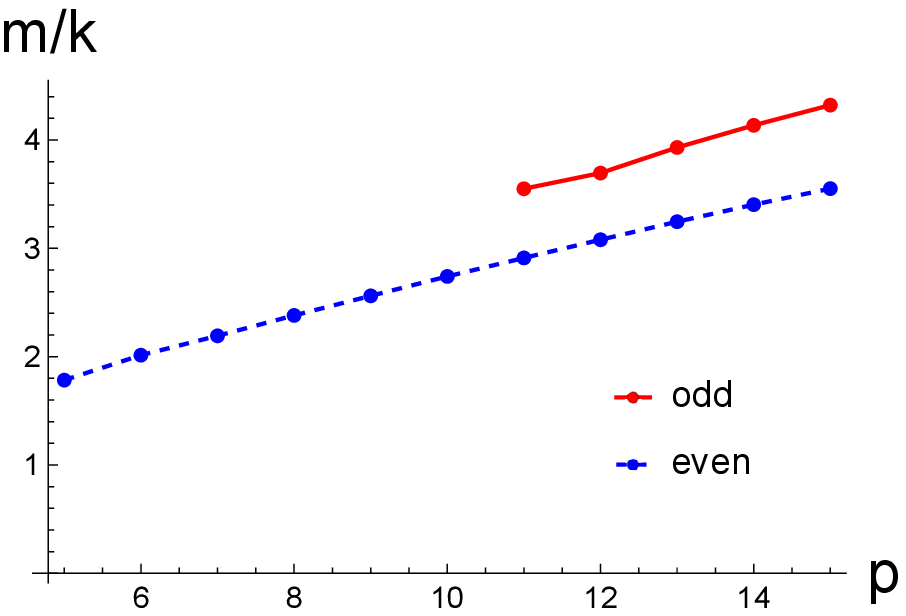}}
    \subfigure[scaled life-time $k\tau$]{\label{figure P 33}
        \includegraphics[width=4.83cm]{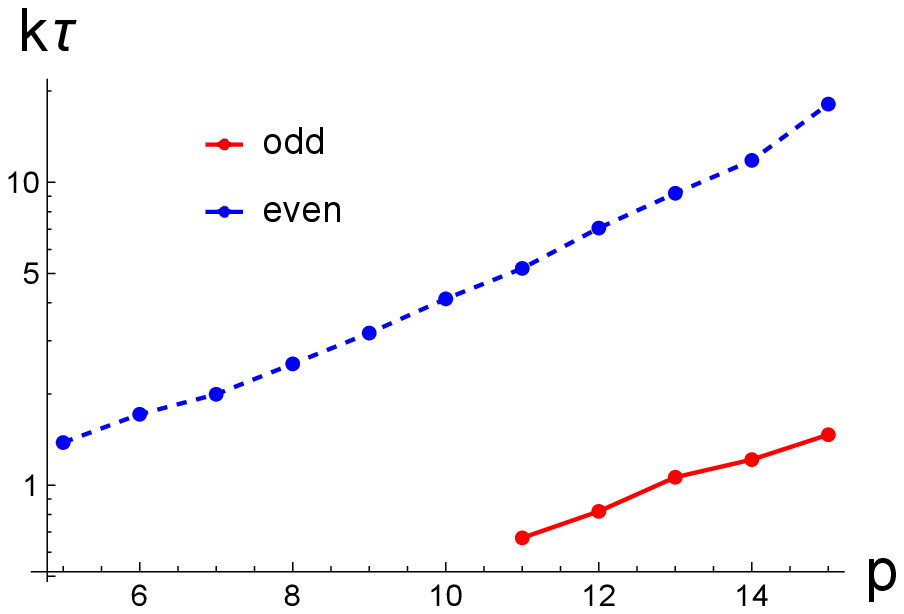}}
    \end{center}
    \caption{The relative probability $P_{max}$, scaled mass ${m}/{k}$,  and scaled life-time $k\tau$  of the resonances of the KR  field for different values of $p$.}  \label{PMLKR}
    \end{figure}

The effective potential of the scalar field  and the relative probability as a function of $m^2/k^2$ are  plotted in Fig.~\ref{VKR} and Fig.~\ref{PKR} for different values of $p$, respectively. And the relative probability, mass, and life-time of the resonances for different values of the parameter $p$ is shown in Fig.~\ref{PMLKR}. It can be seen that the resonances of the KR field behave different with those of the scalar and vector fields. As shown in Fig.~\ref{VKR}, the effective potential possesses a local well only if $p\gtrsim 5$. As shown in the Fig.~\ref{PMLKR}, the first KR resonance appears when $p\gtrsim 5$. Since the zero mode, which has an even parity is not localized on the brane, the first KR resonance has even parity.


\subsubsection{Constraints on the resonances}

It can be seen that the massive resonance of the vector field occurs only if $p\gtrsim 7$, and its mass $m$ is approximately $m/k\simeq 2$ for $p=7$. Considering the experimental constraint on the photon mass, $m_{\gamma}< 10^{-18}\text{eV}$ \cite{Amsler2008}, it leads to $k \lesssim 10^{-18}\text{eV}$ when $p\gtrsim 7$. However, this contradicts the experimental constraint $k\gtrsim 10^{-3}\text{eV}$ form the deviation  from gravitational inverse-square law. Therefore, it implies that no vector resonance exists on the brane and the parameter $p$ is constrained as $0<p\lesssim 7$.

From Fig.~\ref{PMLScalar}, it can be observed that for the resonances of the scalar field, their masses approximate $m \sim \mathcal{O}(k)$, and their lifetimes $\tau$ range from $k^{-1}<\tau<300k^{-1}$. Similarly, as shown in Fig.~\ref{PMLKR}, the mass of KR resonance is roughly $m \sim \mathcal{O}(k)$, and the lifetime $\tau \sim k^{-1}$. For instance, if  $k\sim 10^{-3}\text{eV}$, the masses of the scalar and KR resonances are of order of $10^{-3}\text{eV}$. Correspondingly, the lifetimes of scalar resonances range form $10^{-12} \text{s}$ to $10^{-10}\text{s}$, and the lifetime of KR resonances is roughly $\tau \sim 10^{-12} \text{s}$.

\section{Localization of fermion field}
In this section, we investigate the localization of a spin-$\frac{1}{2}$ fermion field on the brane. The Dirac structure of the fermion fields in the bulk is described by $\tilde{\Gamma}^M=\tilde{e}^{M}_{\bar{M}} \tilde{\Gamma}^{\bar{M}}$ with $\{ \tilde{\Gamma}^M,\tilde{\Gamma}^N \} = 2\tilde{g}^{MN}$. Here $\bar{M},\bar{N},\cdots $ denote the local Lorentz indices and $\tilde{\Gamma}^M$ are the gamma matrices corresponding to the Jordan frame metric  $\tilde{g}^{MN}$. In this set up, it is easy to see that all the equations are the same with the standard case in general relativity except that all the quantities deduced from $\tilde{g}^{MN}$ should be marked by tilde.  With the metric (\ref{tilde g}), we can obtain $\tilde{\Gamma}^M=\text{e}^{-B}\Gamma^M=(\text{e}^{-\tilde{A}}\gamma^{\mu},\text{e}^{-\tilde{A}}\gamma^5)$.
We consider the following action of a massless spin-$\frac{1}{2}$ fermion field coupled to the background chameleon scalar field in a Yukawa-type interaction,
\begin{eqnarray}
\label{fermion action}
   S_f=\int d^5 x \sqrt{-\tilde{g}} \bar{\Psi} \tilde{\Gamma}^M \tilde{D}_M \Psi -\eta \bar{\Psi} f(\phi) \Psi,
\end{eqnarray}
where $f(\phi)$ is a function of the  chameleon scalar field, and the generic covariant derivative $\tilde{D}_M=\pt_M+\tilde{\omega}_M$ is defined with the spin connection $\tilde{\omega}_M= \frac{1}{4} \tilde{\omega}_M^{\bar{M} \bar{N}} \tilde{\Gamma}_{\bar{M}}
\Gamma_{\bar{N}}$, with
\begin{eqnarray}
 \tilde{\omega}_M ^{\bar{M} \bar{N}}
   =&& \frac{1}{2} {\tilde{e}}^{N \bar{M}}(\partial_M \tilde{e}_N ^{\bar{N}}
                      - \partial_N \tilde{e}_M ^{\bar{N}})
   - \frac{1}{2} \tilde{{e}}^{N \bar{N}}(\partial_M \tilde{e}_N ^{\bar{M}}
                      - \partial_N \tilde{e}_M ^{\bar{M}})  \nn \\
   &&- \frac{1}{2} \tilde{{e}}^{P \bar{M}} \tilde{{e}}^{Q \bar{N}} (\partial_P \tilde{e}_{Q
{\bar{R}}} - \partial_Q \tilde{e}_{P {\bar{R}}}) \tilde{{e}}^{\bar{R}} _M.
\end{eqnarray}
Obviously, the fermion field couples to the scalar field not only through the Yukawa interaction, but also  through the determinant $\sqrt{-\tilde{g}}$, the Gamma matrix $\tilde{\Gamma}$ and the generic covariant derivative $\tilde{D}_M$, which are all related to the Jordan frame metric $\tilde{g}_{MN}$.
With the metric (\ref{tilde g}), we obtain the non-vanishing components of the spin connection $\omega_M$,
\begin{eqnarray}
    \label{spin cn}
 \tilde{\omega}_\mu =\frac{1}{2}\tilde{A}'\gamma_{\mu}\gamma_5.
\end{eqnarray}

The variation of (\ref{fermion action}) with respect to $\bar{\Psi}$ yields the five-dimensional Dirac equation,
\begin{eqnarray}
 \label{DIrac5}
 \tilde{\Gamma}^M \tilde{D}_M \Psi=\eta  f(\phi) \Psi,
 \end{eqnarray}
 Substituting Eqs.~(\ref{spin cn}) into the above equation, one can obtain the explicit form of the Dirac equation in the metric (\ref{tilde g})
\begin{eqnarray}
 \left[ \gamma^{\mu}\partial_{\mu}
         + \gamma^5 \left(\partial_z  +2 \tilde{A}' \right)
         -\eta\; \text{e}^{\tilde{A}} f(\phi)
 \right] \Psi =0, \label{DiracEq1}
\end{eqnarray}
Then we employ the chiral decomposition
\begin{equation}
 \Psi(x,z) = \text{e}^{-2\tilde{A}}\sum_n\psi_{Ln}(x) \alpha_{Ln}(z)
 +\sum_n\psi_{Rn}(x) \alpha_{Rn}(z),
\end{equation}
where $\psi_{Ln}(x)=-\gamma^5 \psi_{Ln}(x)$ and
$\psi_{Rn}(x)=\gamma^5 \psi_{Rn}(x)$ are left-handed and
right-handed components of four-dimensional Dirac fields respectively. By substituting  this decomposition into the five-dimensional Dirac equation (\ref{DIrac5}), the four-dimensional Dirac parts $\psi_{Ln}(x)$ and $\psi_{Rn}(x)$ satisfy the four-dimensional Dirac equations
\begin{eqnarray}
 \gamma^{\mu}\partial_{\mu}\psi_{Ln}(x)=m_n\psi_{Rn}(x),	\\
 \gamma^{\mu}\partial_{\mu}\psi_{Rn}(x)=m_n\psi_{Ln}(x).
 \end{eqnarray}
and the five-dimensional Dirac parts $\alpha_{L}(z)$ and $\alpha_{R}(z)$ satisfy the coupled equations
\begin{subequations}
\begin{eqnarray}
 \left[ \partial_z+ \eta\;\text{e}^{\tilde{A}} f(\phi) \right] \alpha_{Ln}(z)
  &=&  ~~m_n \alpha_{Rn}(z), \label{CoupleEq1a}  \\
 \left[ \partial_z- \eta\;\text{e}^{\tilde{A}} f(\phi) \right] \alpha_{Rn}(z)
  &=&  -m_n \alpha_{Ln}(z). \label{CoupleEq1b}
\end{eqnarray}\label{CoupleEq1}
\end{subequations}
After reassembling the two coupled equations, one ultimately achieves the Schr\"odinger-like equations
\begin{eqnarray}
  \label{FermionL}
  \left[-\partial^2_z + V_{L,R}(z) \right]\alpha_{Ln,Rn} &=& m_n^2 \alpha_{Ln,Rn},
\end{eqnarray}
with the effective potentials given by
\begin{eqnarray}
  \label{potentialFermion}
  V_{L,R}(z)&=&\left[\text{e}^{\tilde{A}}\eta f(\phi)\right]^2\mp\pt_z \left[\text{e}^{\tilde{A}}\eta f(\phi)\right].
\end{eqnarray}
Further, to obtain the effective action of a massless and a series of massive four-dimensional fermions on the brane,  the following orthonormal conditions have to be imposed,
\begin{eqnarray}
  \label{orth}
 \int \alpha_{Lm}\alpha_{Ln}&=&\delta_{mn},	\\
  \int \alpha_{Rm}\alpha_{Rn}&=&\delta_{mn},	\\
   \int \alpha_{Lm}\alpha_{Rn}&=&0.	
 \end{eqnarray}
 The chiral zero modes can be solved by applying the factorizing method, given by
 \begin{eqnarray}
  \label{zeroFermion}
 \alpha_{L0,R0}= N_{L0,R0}\text{e}^{\mp\eta\int \text{e}^{\tilde{A}} f(\phi) dz}=\text{e}^{\mp\eta\int \text{e}^{A} b(\phi) f(\phi) dz},
\end{eqnarray}
with $N_{L0,R0}$ the normalization constants. Since we need at least the massless fermion to be localized on the brane, the normalization condition for the zero modes must be satisfied, i.e.,
\begin{eqnarray}
  \label{orth0}
 \int^{+\infty}_{-\infty} \text{e}^{\mp2\eta\int \text{e}^{A} b(\phi) f(\phi) dz} dz<\infty,
 \end{eqnarray}
 where the plus or minus sign is corresponding to the zero mode of left or right-handed four-dimensional fermion. Obviously, only one of the zero modes can be localized on the brane depending on the value of coupling constant $\eta$. 
 
 It is well known that the standard model is a chiral theory, where the left and right-handed fermions transform differently under the electroweak gauge group. To generate the chiral fermions on the brane, one can include a left-handed $SU(2)_L$ doublet $\Psi_L$ and a right-handed singlet  $\Psi_R$ in the five-dimensional bulk, then the corresponding chiral zero modes are picked up by properly setting the coupling constants $\eta_{L/R}$ \cite{PhysRevD.62.084025}. Here, we take the localization of left-handed four-dimensional fermion as an example,
i.e., $ \int^{+\infty}_{-\infty} \text{e}^{-2\eta_L\int \text{e}^{A} b(\phi) f(\phi) dz} dz<\infty$.
Note that the brane is embedded in a five-dimentional asymptotic AdS spacetime, we have $\text{e}^{A(\pm \infty)}=\pm\frac{1}{ kz}$, and the kink configuration of the scalar field yields $\phi(\pm \infty)=\pm \frac{\sqrt{3}\pi}{2} \equiv \pm v_0$.  Therefore,
\begin{eqnarray}
&&\text{e}^{-2\eta_L\int \text{e}^{A} b(\phi) f(\phi) dz} \rightarrow z^{- \frac{2\eta_L b( v_0)f( v_0) }{k}} ~~ \text{as}~~ z\rightarrow + \infty,\\
&&\text{e}^{-2\eta_L\int \text{e}^{A} b(\phi) f(\phi) dz} \rightarrow (- z)^{ \frac{2\eta_L b(- v_0)f(- v_0) }{k}}~~ \text{as}~~ z\rightarrow - \infty.
\end{eqnarray}
Then the normalization condition (\ref{orth0}) is reduced to that:
 $f(\phi)$ must be an odd function of $\phi$, $b(v_0)f(v_0)$ is a non-vanishing constant, and $\eta_L>\frac{k}{b( v_0)f(v_0)}$. As two examples, we can assume that $f(\phi)=\frac{\phi}{\cos^p(\frac{\phi}{\sqrt{3}})}$ or $f(\phi)=\frac{\sin(\phi)}{\cos^p(\frac{\phi}{\sqrt{3}})}$, then the localization of left-handed zero mode is realized for $\eta_L>\frac{k}{v_0}$ or $\eta_L>\frac{k}{\sin v_0}$, respectively. 
 
It is noted that, although the fermion field couples to the chameleon scalar field $\phi$ in several places in the action (\ref{fermion action}), the Yukawa interaction is still essential to localizing the fermion zero mode.
 
%
%

\section{Conclusion}

In this work, we investigated the localization of $q$-form fields and fermion field on the thick brane construct by a chameleon scalar field. We had demonstrated that the localization problem of the vector field can be solved by choosing an appropriate conformal factor $b(\phi)$ in this model. The conditions for localization of various matter fields were obtained. For the $q$-form fields, the zero modes of the scalar and vector fields can be localized, in the  condition that $  b[\phi(\infty)] = z^{r}$
with $r<0$, while the zero mode of KR field can not. Furthermore, the quasi-localization of $q$-form fields was also considered. It was found that the relative probability, mass, and life-time of the resonances increase with the parameter $p$. The parity of first resonances of scalar and vector fields is odd, and the one of the KR field is even. In order to be consistent with both the experimental constraints of the photon mass and the deviation  from gravitational inverse-square law, the parameters were constrained as $0<p\lesssim 7$ and $k \gtrsim  10^{-3}\text{eV}$.

For the fermion field, it couples to the chameleon scalar field not only through the Yukawa interaction, but we found that the term of Yukawa interaction is still necessary in order to localize the fermion zero modes. The condition for localizing the left-handed fermion zero mode was found to be, $f(\phi)$ an odd function of $\phi$, $b(v_0)f(v_0)$ a non-vanishing constant, and $\eta>\frac{k}{b( v_0)f(v_0)}$.

\section*{Acknowledgement}

Y. Zhong acknowledges the support of the Natural Science Foundation of Hunan Province, China (Grant No.~2022JJ40033), the Fundamental Research Funds for the Central Universities (Grants No.~531118010195) and the National Natural Science Foundation of China (No.~12275076). He also thanks the generous hospitality offered by the Center of Theoretical Physics at Lanzhou University where part of this work was completed.  K. Yang acknowledges the support of the National Natural Science Foundation of China under Grant No.~12005174.

%

%

\end{document}